\documentclass{ws-rv9x6}
\usepackage{ws-rv-van}     
\usepackage{graphicx}          
\usepackage{bm}
\newcommand{\mb}[1]{\mathbf{#1}}

\newcommand{\half}{\frac{1}{2}}

\newcommand{\thalf}{\tfrac{1}{2}}

\newcommand{\Tr}{\mbox{Tr}}

\newcommand{\vac}{|0\rangle}
\newcommand{\EFA}{\textsc{efa}}

\newcommand{\be}{\begin{equation}}
\newcommand{\ee}{\end{equation}}
\newcommand{\ba}{\begin{array}}
\newcommand{\ea}{\end{array}}
\newcommand{\bn}{\begin{eqnarray}}
\newcommand{\en}{\end{eqnarray}}
\newcommand{\bt}{\begin{tabular}}
\newcommand{\et}{\end{tabular}}
\newcommand{\bml}{\begin{mathletters}}
\newcommand{\eml}{\end{mathletters}}
\newcommand{\bc}{\begin{center}}
\newcommand{\ec}{\end{center}}
\newcommand{\bi}{\begin{itemize}}
\newcommand{\ei}{\end{itemize}}

\newcommand{\bnll}[1]{\begin{subequations}\label{#1}\begin{eqnarray}}
\newcommand{\enll}{\end{eqnarray}\end{subequations}}
\newcommand{\Bo}{Bogoliubov}

\begin{document}

\chapter[Hartree-Fock-Bogoliubov solution of the pairing Hamiltonian in
finite nuclei]{Hartree-Fock-Bogoliubov solution of the pairing Hamiltonian in
finite nuclei}\label{ra_ch1}

\author[J. Dobaczewski and W. Nazarewicz]
{J. Dobaczewski$^{1,2}$ and W. Nazarewicz$^{1,3,4}$}

\address{
$^{1}${Institute of Theoretical Physics, Faculty of Physics, \\
University of Warsaw, ul. Ho{\.z}a 69, PL-00681 Warsaw, Poland}  \\
$^{2}${Department of Physics, \\ PO Box 35 (YFL), FI-40014
University of Jyv{\"a}skyl{\"a}, Finland} \\
$^{3}${Department of Physics and Astronomy, University of
Tennessee, \\ Knoxville, Tennessee 37996, USA} \\
$^{4}${Physics Division, Oak Ridge National Laboratory, \\ Post
Office Box 2008, Oak Ridge, Tennessee 37831, USA}
}

\begin{abstract}
We present an overview of the Hartree-Fock-Bogoliubov (HFB) theory of nucleonic superfluidity
for finite nuclei. After introducing basic concepts related to pairing
correlations, we show how the correlated pairs are incorporated into the HFB wave function. Thereafter,
we present derivation and structure of the HFB equations within the superfluid nuclear density functional formalism and  discuss several aspects of the theory, including
the unitarity of the Bogoliubov transformation in truncated
single-particle and quasiparticle spaces, form of the pairing
functional, structure of the HFB continuum, regularization and
renormalization of pairing fields, and treatment of pairing in systems with  odd
particle numbers.
\end{abstract}

\body

\section{Introduction}\label{sec1}
Nucleonic pairing  is a ubiquitous  phenomenon underlying many
aspects of  structure and dynamics of atomic nuclei and extended
nuclear matter\cite{[Dea03],(Bri05a)}. The crucial role of nucleonic
superfluidity lies in its emergent nature. Indeed, while the
correlation energy due to pairing is a small correction to the
nuclear binding energy, the superfluid  wave function represents an
entirely different phase described by new quasiparticle degrees of
freedom. The road to this new phase is associated with a phase
transition  connected with a symmetry breaking, and this underpins
the nonperturbative nature of pairing.

Many facets of nucleonic superfluidity -- including those related to
phenomenology and theory of pairing -- are discussed in this volume
\cite{(BroZel)}. Here, we outline several aspects of nucleonic
superfluidity  within the  framework of the nuclear density
functional theory (DFT). The main building blocks of nuclear DFT are
the effective mean fields, often represented by local nucleonic
densities and currents. When compared to the electronic  DFT for the
superconducting state\cite{[deG66a],(Koh89),(Kur99)} the unique
features of the nuclear variant are (i) the presence of two kinds of
fermions, protons and neutrons, (ii) the absence of external
potential, and (iii) the need for symmetry restoration in a finite
self-bound system. In the context of pairing, nuclear superfluid  DFT
is a natural extension of the traditional BCS theory for electrons
\cite{[Bar57]} and nucleons\cite{[Boh58]}, and a  tool of choice for
describing complex, open-shell  nuclei.

At the heart of nuclear  DFT lies  the energy density functional (EDF). The
requirement that the total energy be minimal under a variation of the
densities leads to the Hartree-Fock-Bogoliubov (HFB; or Bogoliubov-de
Gennes) equations.  The quasiparticle vacuum associated with the HFB
solution is a highly correlated state that allows a simple
interpretation of various phenomena it the language of pairing mean
fields and associated order parameters.

This paper is organized as follows. Section \ref{sec2} describes the
essentials of the general pair-condensate state. The HFB theory is
outlined in Sec.~\ref{sec3}. The Bogoliubov sea, related to the
quasiparticle-quasihole symmetry of  the HFB Hamiltonian is discussed
in Sec.~\ref{sec4} and Sec.~\ref{sec10} is devoted to the form of the
nuclear pairing EDF. The quasiparticle energy spectrum of HFB
contains both discrete bound states and continuum unbound states. The
properties of the associated quasiparticle continuum are reviewed in
Sec.~\ref{sec5}. Section~\ref{sec7} describes the extension of the
HFB formalism to odd-particle systems and quasiparticle blocking.
Finally, conclusions are contained in Sec.~\ref{sec11}.

\section{Basics of pairing correlations}\label{sec2}
In quantum mechanics of finite many-fermion systems, pairing
correlations are best described  in terms of  number
operators $\hat{N}_\mu=a^+_\mu a_\mu$, where $\mu$ represents any
suitable set of single-particle (s.p.) quantum numbers. Thus we may have,
e.g., $\mu\equiv\bm{k}\sigma$ for plane waves of spin-$\half$
particles (electrons); $\mu\equiv\bm{r}\sigma\tau$ for  spin-$\half$ and
isospin-$\half$ nucleons localized in space at position $\bm{r}$; and
$\mu\equiv n,\ell,j,m$ for fermions moving in a spherical potential
well. Since $\hat{N}^2_\mu=\hat{N}_\mu$, the number operators are
projective; hence, one can -- at least in principle  -- devise
an experiment that would project any quantum many-fermion state
$|\Psi\rangle$ into its component with exactly one fermion occupying
state $\mu$. As the rules of quantum mechanics stipulate, any such
individual measurement can only give 0 or 1 (these are the
eigenvalues of $\hat{N}_\mu$), whereas performing such measurements
many times, one could experimentally determine the
occupation probabilities $v^2_\mu=\langle\Psi|\hat{N}_\mu|\Psi\rangle$.

Along such lines, we can  devise an
experiment that would determine the {\em simultaneous} presence of two
fermions in  different orthogonal s.p. states $\mu$ and
$\nu$. Since the corresponding number operators $\hat{N}_\mu$ and
$\hat{N}_\nu$ commute, one can legitimately ask quantum
mechanical questions about one-particle occupation probabilities
$v^2_\mu$ and $v^2_\nu$, as well as about the two-particle occupation.
The latter one reflects the   simultaneous presence of two fermions in state
$|\Psi\rangle$:
$v^2_{\mu\nu}=\langle\Psi|\hat{N}_\mu\hat{N}_\nu|\Psi\rangle$.
In this way, one can experimentally determine the pairing correlation
between states $\mu$ and $\nu$ as the {\em excess} probability
\begin{equation}\label{eq:1}
P_{\mu\nu} = v^2_{\mu\nu} -  v^2_{\mu}v^2_{\nu} ,
\end{equation}
of finding two fermions simultaneously over that of finding them in
independent, or sequential, measurements. Such a definition of
pairing is independent of its coherence, collectivity, nature of
quasiparticles, symmetry breaking, thermodynamic limit, or many other
notions that are often associated with the phenomenon of pairing. In
terms of occupations, pairing can be viewed as  a  measurable
property of any quantum many-fermion state.

Obviously, no pairing correlations are present  in a quantum state
that is an eigenstate of $\hat{N}_\mu$ or $\hat{N}_\nu$,
such as the Slater determinant.
The beauty of the BCS ansatz is in providing us
with a model $N$-fermion state, in which pairing correlations are
explicitly incorporated:
\begin{equation}\label{eq:2}
|\Phi_N\rangle = {\cal{}N}_N\left(\sum_{\mu>0}s_\mu z_\mu a^+_{\tilde\mu} a^+_{\mu}
\right)^{N/2}|0\rangle ,
\end{equation}
where the summation $\mu>0$ runs over the representatives of pairs
($\tilde\mu,\mu$) of s.p.\  states (that is, any one state of
the pair is included in the sum, but not both), $z_{\tilde\mu} = z_\mu$
are real positive numbers, $s_{\tilde\mu} = - s_\mu$ are arbitrary
complex phase factors, and ${\cal{}N}_N$ is the overall normalization
factor.

It  now becomes a matter of technical convenience to employ
a particle-number mixed state,
\begin{equation}\label{eq:3}
|\Phi\rangle = {\cal{}N}\sum_{N=0,2,4,\ldots}^\infty\frac{|\Phi_N\rangle}{{\cal{}N}_N(N/2)!}
= {\cal{}N}\exp\left(\sum_{\mu>0}s_\mu z_\mu a^+_{\tilde\mu} a^+_{\mu}
\right)|0\rangle,
\end{equation}
in which the pairing correlations (\ref{eq:1}) are:
\begin{equation}\label{eq:4}
P_{\mu\nu} = v^2_{\mu}u^2_{\nu}\delta_{{\tilde\mu}\nu}
\quad\mbox{for}\quad v^2_{\mu}=\frac{z^2_{\mu}}{1+z^2_{\mu}}
\quad\mbox{and}\quad u^2_{\nu}=\frac{1        }{1+z^2_{\nu}}.
\end{equation}
In terms of the s.p. occupations, the state $|\Phi\rangle$ assumes
the standard BCS form:
\begin{equation}
|\Phi\rangle= \prod_{\mu>0}\left(u_\mu + s_\mu v_\mu a^+_{\tilde\mu}
a^+_{\mu}\right)|0\rangle.
\end{equation}
In this many-fermion state, the s.p.\ states $\tilde\mu$ and $\mu$
are paired, that is, $|\Phi\rangle$ can be viewed as a
pair-condensate. For $z_{\mu}$=1, the pairing correlation
$P_{{\tilde\mu}\mu}$ (\ref{eq:1}) equals 1/4; in fact, in this state,
it is twice more likely to find a pair of fermions
($v^2_{{\tilde\mu}\mu}$=1/2) than to find these two fermions
independently ($v^2_{{\tilde\mu}}v^2_{\mu}$=1/4). For particle-number
conserving states (\ref{eq:2}), the occupation numbers can be
calculated numerically; qualitatively the results are fairly similar,
especially for large numbers of particles.

At this point, we note that the most general pair-condensate state (\ref{eq:3})
has the form of the Thouless state,
\begin{equation}\label{eq:5}
|\Phi\rangle
= {\cal{}N}\exp\left(\thalf\sum_{\nu\mu}Z^*_{\nu\mu} a^+_{\nu} a^+_{\mu}
\right)|0\rangle ,
\end{equation}
in which pairs ($\tilde\mu,\mu$) do not appear explicitly. However,
there always exists a unitary transformation $U_0$ of the antisymmetric
matrix $Z$ that brings it to the canonical form
$(U_0^+Z^*U_0^*)_{\nu\mu}= s_{\mu}z_{\mu}\delta_{{\tilde\mu}\nu}$ (the
Bloch-Messiah-Zumino theorem\cite{[Blo62],[Zum62]}). Therefore,
pairs are present in any arbitrary Thouless state (the so-called
canonical pairs), and they can be made explicitly visible by a simple
basis transformation.

The canonical pairs exist independently of any
symmetry of the Thouless state. In the particular case of a
time-reversal-symmetric state, $\hat{T}|\Phi\rangle=|\Phi\rangle$,
they can be associated with the time-reversed s.p.\ states,
$\tilde{\mu}\equiv\bar{\mu}$. The ground-states of even-even nuclei
can be described in this manner. However, the appearance of pairing
phase  does not hinge on this particular symmetry -- states in
rotating nuclei (in which time-reversal symmetry is manifestly
broken) can also be paired. In this latter case the canonical states
are less useful, because they cannot be directly associated with the
eigenstates of the HFB Hamiltonian.

\section{Hartree-Fock-Bogoliubov theory}\label{sec3}

The simplest route to the HFB theory is to employ
the variational principle to a two-body Hamiltonian using
Thouless states (\ref{eq:5}) as trial wave functions.
The variation of the average energy with respect to the antisymmetric matrix $Z$
results in the HFB equation in the matrix representation,
${\cal{}H}{\cal{}U}={\cal{}U}{\cal{}E}$, or explicitly,
\begin{equation}
\label{eq7.154}
\left(\ba{cc} T+\Gamma & \Delta \\
               -\Delta^*& -{T}^*-\Gamma^*\ea\right)
\left(\ba{cc} U & V^* \\
              V & U^*  \ea\right) =
\left(\ba{cc} U & V^* \\
              V & U^*  \ea\right)
\left(\ba{cc} E &     0 \\
                0 &  -E  \ea\right) ,
\end{equation}
where $T_{\mu\nu}$ is the matrix of the one-body kinetic energy,
$\Gamma_{\mu\nu}=\sum_{\mu'\nu'}V_{\mu\mu';\nu\nu'}\rho_{\nu'\mu'}$
and
$\Delta_{\mu\mu'}=\thalf\sum_{\nu\nu'}V_{\mu\mu';\nu\nu'}\kappa_{\nu\nu'}$
are the so-called particle-hole and particle-particle mean fields,
respectively,
obtained by averaging two-body matrix elements $V_{\mu\mu';\nu\nu'}$
with respect to the density matrix
$\rho_{\nu'\mu'}=\langle\Phi|a^+_{\mu'}a_{\nu'}|\Phi\rangle$ and
pairing tensor
$\kappa_{\nu\nu'}=\langle\Phi|a_{\nu'}a_{\nu}|\Phi\rangle$, and $E$
is the diagonal matrix of quasiparticle energies\cite{[RS80]}.

The matrices ${\cal{}H}$ and ${\cal{}U}$ are referred to as the HFB Hamiltonian
and {\Bo} transformation, respectively, and columns of ${\cal{}U}$
(eigenstates of ${\cal{}H}$) are vectors of  quasiparticle states. The HFB
equation (\ref{eq7.154}) possesses the quasiparticle-quasihole
symmetry. Namely, for each quasiparticle state
$\chi_\alpha$ (the
$\alpha$-th column of ${\cal{}U}$) and energy $E_\alpha$ there exists
a quasihole state
$\phi_\alpha$
of opposite energy $-E_\alpha$,
\begin{equation}
\label{eq:9}
\chi_\alpha=\left(\ba{c}U_{\mu\alpha}\\V_{\mu\alpha}\ea\right) , \quad
\phi_\alpha=\left(\ba{c}V^*_{\mu\alpha}\\U^*_{\mu\alpha}\ea\right)  .
\end{equation}
That is, the spectrum of ${\cal{}H}$ is
composed  of pairs of states with opposite energies. In most cases,
the lowest total energy is obtained by using the eigenstates
with $E_\alpha>0$ as quasiparticles $\chi_\alpha$ and those with
$E_\alpha<0$ as quasiholes $\phi_\alpha$, that is, by occupying the
negative-energy eigenstates. States $\chi_\alpha$ and $\phi_\alpha$
can usually be related through a self-consistent discrete symmetry,
such as  time reversal, signature, or
simplex.\cite{[Goo74],[Dob00a],[Fra00]}.

The HFB equation (\ref{eq7.154}) is also valid in a more general
case, when the total energy is not equal to the average of any many-body Hamiltonian.
Within the DFT, it stems from the
minimization of the binding energy given by an EDF ${\cal{}E}(\rho,\kappa,\kappa^*)$,
subject to the condition of the generalized density
matrix being projective, that is, ${\cal{}R}^2={\cal{}R}$ for
\be\label{eq4.72}
     {\cal{}R}   =     \left(\ba{cc} \rho     & \kappa \\
                                   -\kappa^* & 1-\rho^*   \ea\right)
               = \left(\ba{cc} V^*V^T & V^*U^T \\
                               U^*V^T & U^*U^T   \ea\right)
               = \left(\ba{c}  V^*   \\ U^*      \ea\right)
                 \raisebox{+0.5em}{$
                 \left(\ba{cc} V^T    & U^T      \ea\right)$}
= \sum_\alpha \phi_\alpha\phi_\alpha^+ .
\ee
In this case, the mean fields are obtained as functional derivatives of EDF:
$\Gamma_{\mu\nu}=\partial{\cal{}E}/\partial\rho_{\nu\mu}$
and
$\Delta_{\mu\mu'}=\partial{\cal{}E}/\partial\kappa^*_{\mu\mu'}$. As
is the case in DFT, densities (here the
density matrix and pairing tensor) become the fundamental degrees of
freedom, whereas the state $|\Phi\rangle$ acquires the
meaning of an auxiliary entity (the Kohn-Sham
state\cite{[Koh65a]}). Indeed, for any arbitrary generalized density
matrix ${\cal{}R}$ (\ref{eq4.72}), one can always find the
corresponding state $|\Phi\rangle$. For that, one determines the
{\Bo} transformation ${\cal{}U}$ as the matrix of its eigenvectors,
${\cal{}R}{\cal{}U}={\cal{}U}\left(\ba{cc}0&0\\0&1\ea\right)$;
 the Thouless state $|\Phi\rangle$ (\ref{eq:5}) corresponds to
$Z=VU^{-1}$. Consequently, the paired state $|\Phi\rangle$ of DFT is not
interpreted as a wave function of the system -- it only
serves as a model  for  determining  one-body densities. Nonetheless,
these  densities {\em are}  interpreted as those associated with the
(unknown) exact eigenstate of the system.

Unrestricted variations of the EDF are not meaningful.
Indeed, since Thouless states (\ref{eq:3}) are mixtures of components
with different particle numbers, absolute minima will usually  correspond to
average particle numbers that are unrelated to those one would like
to describe.
In particular, for self-bound systems governed by attractive two-body
forces (nuclei), by adding more and more particles one could infinitely
decrease the total energy of the system. Therefore, only constrained variations
make sense, that is, one has to minimize not the total energy
${\cal{}E}(\rho,\kappa,\kappa^*)$, but the so-called Routhian,
${\cal{}E}'(\rho,\kappa,\kappa^*)={\cal{}E}(\rho,\kappa,\kappa^*)+{\cal{}C}(\rho)$,
where ${\cal{}C}$ is a suitably chosen penalty functional,
ensuring that the minimum appears at prescribed average values
of one-body operators. In particular, the average total number of
particles can be constrained by
${\cal{}C}(\rho)=-\lambda\langle\Phi|\hat{N}|\Phi\rangle=-\lambda\Tr(\rho)$
(linear constraint) or ${\cal{}C}(\rho)=C_N[\Tr(\rho)-N_0]^2$ (quadratic
constraint),\cite{[Flo73],[Sta10]} where $\lambda$ becomes the Fermi
energy corresponding to $N_0$ fermions.

For different systems and for different applications, various
constraints ${\cal{}C}(\rho)$ can be implemented; for example, in
nuclei one can simultaneously constrain numbers of protons and
neutrons, as well as multipole moments of matter or charge
distributions. When the total energy is a concave function of
relevant one-body average values, quadratic constraints are
mandatory\cite{[Flo73],[Sta10]}. The minimization of
${\cal{}E}'(\rho,\kappa,\kappa^*)$ requires solving the HFB equation
for the quasiparticle Routhian ${\cal{}H}'$, which, for the simplest
case of the constraint on the total particle number, reads
${\cal{}H}'={\cal{}H}-\lambda\left(\ba{cc}1&0\\0&-1\ea\right)$.

\begin{figure}[h]
{\centerline{\includegraphics[width=0.50\textwidth]{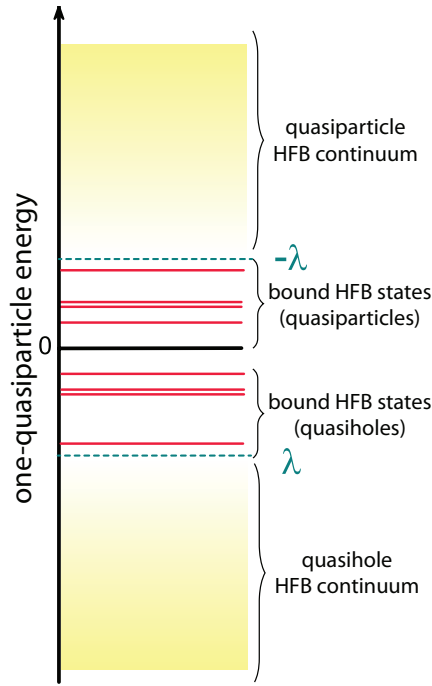}}}
\caption{Quasiparticle spectrum of the HFB Hamiltonian.
The bound states exist  in the
energy region $|E_i|\leqslant-\lambda$, where $\lambda$ is the chemical potential (always negative for a particle-bound system).\label{Sea}}
\end{figure}

Finally, let us mention that in the coordinate space-spin(-isospin)
representation, the HFB equation (\ref{eq7.154}) acquires
particularly interesting form, which in condensed matter and atomic
literature is called Bogoliubov-de Gennes equation\cite{[deG66a]}.
In the coordinate representation,  quasiparticle vectors become
two-component wave functions, which -- in finite systems -- acquire
specific asymptotic properties\cite{[Bul80],[Dob84],[Bel87],[Dob96]}
determining the asymptotic behavior of local densities.
The quasiparticle energy spectrum of HFB contains discrete bound
states, resonances, and non-resonant continuum states. As illustrated
in Fig.~\ref{Sea}, the bound HFB solutions exist only in the
energy region $|E_i|\leqslant-\lambda$. The quasiparticle continuum
with $|E_i| > -\lambda$ consists of non-resonant continuum and
quasiparticle resonances, see Sec.~\ref{sec6}.

\section{Beneath the bottom of the Bogoliubov sea}\label{sec4}

Because of the quasiparticle-quasihole symmetry (Sec.~\ref{sec3}),
the spectrum of the HFB Hamiltonian contains  as many negative as
positive eigenvalues. Therefore, the HFB equation (\ref{eq7.154})
constitutes an eigenvalue problem for the operator unbounded from
below, and the HFB spectrum extends from minus to plus infinity,
see Fig.~\ref{Sea}.
Moreover, one-body densities (density matrix and pairing tensor)
are given by infinite sums over all negative-energy (quasihole) states,
cf.~Eq.~(\ref{eq4.72}). In analogy to the Fermi sea of occupied
states, which appears in the Hartree-Fock (HF) theory, we call the
set of quasihole states the  {\Bo} sea. Note again that the Fermi sea
extends over a finite range of energies -- from the bottom of the HF
potential up to the Fermi energy -- whereas the {\Bo} sea is
infinitely deep, in a nice analogy with the sea of one-electron
states of the relativistic Dirac equation.

In practice, since infinite sums over the {\Bo} sea cannot be carried out,
the number of pairing-active states must be truncated.
Two different ways of achieving this goal are most often
implemented, namely, solution of the HFB equations in a finite
s.p.\  space (e.g., the so-called two-basis
method\cite{[Gal94]}) and truncation of the summation in the
quasiparticle space. The second method correspond to creating
an artificial finite bottom of the {\Bo} see. In this section we
discuss the consequences of neglecting the
quasihole states that are below the bottom of that sea\cite{[Dob05f],[Bor06b]}.

The main problem concerns the calculation of the pairing tensor,
which is the sum of products of upper and lower components of
quasihole states, cf.~Eq.~(\ref{eq4.72}). When this sum is performed
over the infinite complete set of quasiparticle states, the resulting
pairing tensor is antisymmetric, while for truncated sums it may
acquire a symmetric part. Usually the symmetric component is
small;\cite{[Dob05f]} hence, can be neglected. However, its very
existence means that the many-fermion state which would have had such
a pairing tensor simply does not exist. The smallness of the
symmetric part can be deceiving, because the symmetric pairing tensor
corresponds to a many-boson system.  Consequently,
 appearance of the symmetric component implies the violation of the Pauli principle. This is a
potentially dangerous situation -- within a variational theory one
should avoid the boson sector whose ground state is way
below the fermionic ground state.

A solution to this problem\cite{[Dob05f]}, discussed below, consists in
marrying the two truncation methods mentioned above.  That is, we shall first use the
quasiparticle truncation method to define the appropriate
s.p.\  cutoff, and then the HFB equations are solved
in this truncated space, leading to a perfectly
antisymmetric pairing tensor.

Let us consider the case of truncated summations over the {\Bo} sea
and assume that we have kept only $K$ quasihole states. In order to
maintain the quasiparticle-quasihole symmetry,  we apply the same truncation to quasiparticle  and quasihole states, that
is, we also keep $K$ quasiparticle partner states. This is convenient, and
always possible, because the quasiparticle (unoccupied) states do not
impact HFB densities. Then, matrix ${\cal{}U}$ becomes
rectangular -- it has less columns ($2K$) than rows. Since all kept
quasiparticle and quasihole states are orthonormal, we still have
${\cal{}U}^+{\cal{}U} = 1$. However, since now the quasiparticle
space is not complete, ${\cal{}U}$ is not anymore unitary, and the
product ${\cal{}U}{\cal{}U}^+ = {\cal{}P}$ is not equal to unity. All
what remains is the  hermitian and projective property of  ${\cal{}P}$:
\be\label{eq-a.1}
{\cal{}P}^+={\cal{}P}, \quad
{\cal{}P}^2={\cal{}P}.
\ee
Since
$\Tr{\cal{}P}=2K$, ${\cal{}P}$ has exactly $2K$ eigenvalues
equal to 1.

In its explicit form, matrix ${\cal{}P}$ reads:
\be\label{eq-a.2}
{\cal{}P}
 =   \left(\ba{cc} P   & Q \\
                   Q^* & P^* \ea\right)
 =   \left(\ba{cc} UU^+ + V^*V^T,  & UV^+ + V^*U^T \\
                   VU^+ + U^*V^T,  & VV^+ + U^*U^T \ea\right).
\ee
In terms of $P$ and $Q$, Eqs.~(\ref{eq-a.1}) can be written as:
\be\label{eq-a.3}
P^+ = P, \quad Q^T = Q \quad\mbox{and}\quad
P^2 + QQ^+ = P, \quad
PQ  + QP^* = Q,
\ee
where $\Tr{P}=K$.

Properties of matrices $P$ and $Q$ can be most easily discussed in
the particle basis that diagonalizes $P$. Suppose that column $f$ is
an eigenvector of $P$ with eigenvalue $p$, that is, $Pf = pf$. From
Eqs.~(\ref{eq-a.3}) it follows that $p$ must be between 0
and 1. Moreover, if $Qf^*$ is not equal to zero, then $Qf^*$ is an
eigenvector of $P$ with eigenvalue 1$-$$p$, that is, P$(Qf^*) =
(1-p)Qf^*$. Conversely, if $Qf^*$ is equal to zero, than $p$=0 or
$p$=1.

Altogether, the spectrum of $P$ can be divided into three regions:
(i) $i$ states with $p_\nu$=1,
where all matrix elements of $Q$ vanish,  $Q_{\nu\nu'}$=0, (ii) $2k$ states with
0$<$$p_\nu$$<$1, where eigenvectors are arranged in pairs
$p_{\tilde{\nu}}$=1$-$$p_{\nu}$ such that the only non-vanishing matrix
elements of $Q$ are
\be\label{eq-a.7}
Q_{\nu\tilde{\nu}} = Q_{\tilde{\nu}\nu} = q_{\nu} = \sqrt{p_{\nu}(1-p_{\nu})},
\ee
and (iii) states with $p_\nu$=0, where again all matrix elements of $Q$ vanish.
In practical calculations of solving the HFB equation in infinite-dimensional
quasiparticle spaces (like the coordinate representation),
the first region almost never appears ($i=0$), and then $k=K$. However, for
truncated quasiparticle spaces, the third region always exists and
contains the null space of $P$.

We now see that when $K$ quasiparticles are included in the quasiparticle
space and $i=0$, in the particle space there appears a basis of 2$K$
s.p.\  states, which we call natural basis. Each state in
the first half of the natural basis has its partner in the second
half.
By ordering the eigenvalues of $P$
and neglecting the zero
eigenvalues for $\nu>2K$, we can write
matrices $P$ and $Q$ in a general form:
\newlength{\di}
\setlength{\di}{0.1em}
\be
P=\left(
\ba{c@{\hspace{\di}}c@{\hspace{\di}}c@{\hspace{\di}}c@{\hspace{\di}}|c@{\hspace{\di}}c@{\hspace{\di}}c@{\hspace{\di}}c}
 p_{1} & 0     & \ldots & 0     & 0            & 0           & \ldots & 0           \\
 0     & p_{2} & \ldots & 0     & 0            & 0           & \ldots & 0           \\
\multicolumn{4}{c|}{\dotfill} & \multicolumn{4}{c}{\dotfill}                         \\
 0     & 0     & \ldots & p_{K} & 0            & 0           & \ldots & 0           \\
\hline
 0     & 0     & \ldots & 0     & 1$$-$$p_{K}  & 0           & \ldots & 0           \\
\multicolumn{4}{c|}{\dotfill} & \multicolumn{4}{c}{\dotfill}                         \\
 0     & 0     & \ldots & 0     & 0            & 1$$-$$p_{2} & \ldots & 0           \\
 0     & 0     & \ldots & 0     & 0            & 0           & \ldots & 1$$-$$p_{1} \\
\ea
\right), \quad
Q=\left(
\ba{c@{\hspace{\di}}c@{\hspace{\di}}c@{\hspace{\di}}c@{\hspace{\di}}|c@{\hspace{\di}}c@{\hspace{\di}}c@{\hspace{\di}}c}
 0     & 0     & \ldots & 0     & 0            & 0           & \ldots & q_1         \\
 0     & 0     & \ldots & 0     & 0            & q_2         & \ldots & 0           \\
\multicolumn{4}{c|}{\dotfill} & \multicolumn{4}{c}{\dotfill}                         \\
 0     & 0     & \ldots & 0     & q_K          & 0           & \ldots & 0           \\
\hline
 0     & 0     & \ldots & q_K   & 0            & 0           & \ldots & 0           \\
\multicolumn{4}{c|}{\dotfill} & \multicolumn{4}{c}{\dotfill}                         \\
 0     & q_2   & \ldots & 0     & 0            & 0           & \ldots & 0           \\
 q_1   & 0     & \ldots & 0     & 0            & 0           & \ldots & 0           \\
\ea
\right).
\ee
If there appear $i>0$ states with $p_\nu=1$,
the number of paired states decreases to $2k=2K-2i$ and the
size of the natural basis decreases to $i+2k=2K-i$.

The HFB equations can now be solved in the finite natural basis,
whereupon the pairing tensor becomes exactly antisymmetric and the
dangerous violations of the Pauli principle are removed exactly\cite{[Dob05f]}.
The advantage of this method is in the fact that the truncated s.p.\
space is not arbitrarily cut but it is adjusted to the truncated
quasiparticle space.

\section{Pairing functional}\label{sec10}

The form of the most general pairing EDF that is quadratic in local
isoscalar and isovector densities has been discussed  in
Refs.\cite{[Per04],[Roh10]}. Because of  the lack of nuclear
observables that can constrain coupling constants of this general
pairing functional, current realizations are much simpler. A commonly
used effective pairing interaction is  the zero-range pairing force
with the density-dependent form factor
\cite{[Boc67],[Cha76],[Kad78],[Ter95],[Dob01a]}:
\begin{equation}\label{DDP}
f_{\text{pair}}(\bm{r}) =
V_0\left\{1+x_0\hat{P}^\sigma
-\left[\eta \frac{\rho_0(\bm{r})}{\rho_c}\right]^\alpha
(1+x_3\hat{P}^\sigma)\right\},
\end{equation}
where  $\hat{P}^\sigma$ is the usual spin-exchange operator and
$\rho_0$=0.16\,fm$^{-3}$. When only the isovector pairing is studied,
the exchange parameters $x_0$ and $x_3$ are usually set to 0.
However, in the general case of coexisting isoscalar and isovector
pairing correlations, nonzero values of $x_0$ and $x_3$ have to be
used. Pairing interactions corresponding to $\eta$=0, 0.5, and 1,
are usually referred to as volume, mixed, and surface pairing,
respectively\cite{[Dob01c],[Dob02c],[Dug01w]}. The volume pairing
interaction  acts primarily inside the nuclear volume while the
surface pairing generates pairing fields peaked around or outside
the  nuclear surface.

Another form of density dependence has been
suggested in Ref.\cite{[Fay96]} and successfully applied
\cite{[Fay00]} to explain odd-even effects in charge radii. As
discussed in Refs.\cite{[Dob01a],[Rot09a]}, different assumptions
about the density dependence may result in  differences of pairing
fields in very neutron rich  nuclei. However, the results of the
global survey\cite{[Ber09a]} suggest that -- albeit  there is a
slight favoring of the surface interaction  -- one cannot reliably
extract the density dependence of the effective pairing interaction
(\ref{DDP}) from the currently available experimental odd-even mass
differences, limited to nuclei with a modest neutron excesses (see
also Refs.\cite{[Dob01],[Dug01w],[San05]}).

A timely question, related to the density dependence, is whether there
is an effective isospin dependence of the  pairing interaction. The
global survey\cite{[Ber09a]} of odd-even staggering of binding
energy indicates that the effective pairing strength $V_0$ for
protons is larger than for neutrons, and the recent large-scale
optimizations of the nuclear EDF are consistent with this finding
\cite{[Kor10b],[Kor12]}. This can be attributed to the
isospin-dependent contribution to pairing  from the Coulomb
interaction\cite{[Ang01a],[Les08],[Nak11]} or to induced pairing due
to the coupling to collective  excitations\cite{[Bar99],[Ter02a]}.
To account for those effects, an extended density dependence has been
proposed\cite{[Mar07],[Mar08],[Yam09]} that involves the local
isovector density $\rho_1(\bm{r})$.

Little is known about the isoscalar pairing functional. The local
isovector pairing potential\cite{[Per04],[Roh10]} $\vec{\breve{U}}(\bm{r})$
is proportional to the isovector pair density $\vec{\breve{\rho}}$
whereas the isoscalar {pairing} potential $\breve{\bm{\Sigma}}_0(\bm{r})$
is a vector proportional to the isoscalar-vector  {pairing} spin density
$\breve{\bm{s}}_0$.
Then, the isoscalar pairing field,
\begin{equation}\label{isoscmf}
\breve{h}_0(\bm{r})
= \breve{\bm{\Sigma}}_0\cdot\hat{\bm{\sigma}} \propto
\breve{\bm{s}}_0\cdot\hat{\bm{\sigma}}
\end{equation}
is the projection of the quasiparticle's spin on the
proton-neutron pairing field.
Physically, $\vec{\breve{\rho}}$ represents the
density of $S$=0, neutron-neutron, proton-proton, and proton-neutron
pairs, whereas the vector field $\breve{\bm{s}}_0$  describes  the spin
distribution of $S$=1 pn pairs (that is, it contains all magnetic
components of $S$=1 pn pairing field).

\begin{figure}[htb]
\centerline{\includegraphics[width=0.55\textwidth]{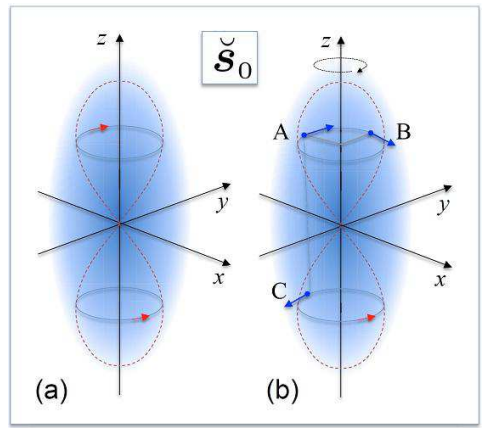}}
\caption{(a) Schematic illustration of the isoscalar vector pairing
field $\breve{\bm{s}}_0$ in the case of conserved axial and mirror
symmetries\cite{[Roh10]}. The field is solenoidal, with vanishing
third component. (b) Under rotation around the third (symmetry) axis,
the field at the point  $\bm{r}_A$ is transformed to the position
$\bm{r}_B$. Likewise, under ${\cal{}R}_x$, the field is transformed
to   $\bm{r}_C$. While neither of these operations leave the
individual vector $\breve{\bm{s}}_0(\bm{r}_A)$ invariant, the field
as whole does not change.
}
\label{pnpair}
\end{figure}

Symmetries of the isoscalar pairing mean-fields have been studied in
detail in Ref.\cite{[Roh10]}. As an example,  lines of the
solenoidal field $\breve{\bm{s}}_0$ -- present in the generalized
pairing theory that mixes proton and neutron orbits -- are
schematically shown in Fig.~\ref{pnpair}. It is interesting to note
that for the geometry of  Fig.~\ref{pnpair},  the third component
$\breve{\bm{s}}_{0z}$, associated with the $M$=0 isoscalar pairing
field vanishes. That is, the solenoidal pairing field is created by
the two components with $M$=$\pm 1$. One can thus conclude that the
assumption of axial symmetry, or signature, does not preclude the
existence of isoscalar pairing.

\section{Pairing and the HFB continuum}\label{sec6}

The structure of  HFB continuum indicated schematically in Fig.~\ref{Sea} has been a
subject of many works
\cite{[Bul80],[Dob84],[Bel87],[Dob96],[Fay98b],[Gra01],[Gra02],[Bor06],[Oba09],[Zha11],[Pei11]}. Within the real-energy HFB framework, the proper
theoretical treatment of the HFB continuum is fairly  sophisticated
since the scattering boundary conditions must be met. One way of
tackling this problem is the coordinate-space Green's function
technique\cite{[Bel87],[Oba09],[Zha11]}. If the outgoing boundary
conditions are imposed, the  unbound HFB eigenstates have complex
energies; their imaginary parts are related to the particle
width\cite{[Bel87]}. The complex-energy spherical HFB problem has
been formulated and implemented   within the Gamow HFB (GHFB)
approach of Ref.\cite{[Mic08]}.

In addition to the methods that  directly employ proper asymptotic
boundary conditions for  unbound HFB eigenstates, the quasiparticle
continuum of HFB can be approximately treated by means of a
discretization method.  The commonly used approach is to impose  the
box boundary conditions in the coordinate-space calculations
\cite{[Dob84],[Dob96],[Obe03],[Ter03],[Yam05a],[Pei08a],[Pei11]}, in
which wave functions are spanned by a basis of orthonormal functions
defined on a lattice in  coordinate space and enforced to be zero at
box boundaries. In this way, referred to as the  $\mathcal{L}^2$
discretization, quasiparticle continuum of HFB  is represented by a
finite number of box  states. It has been demonstrated by explicit
calculations for   weakly bound nuclei\cite{[Mic08],[Gra01],[Pei11]}
that such a box discretization is accurate  when compared to the
exact results. Alternatively, diagonalizing the HFB matrix in the
P\"oschl-Teller-Ginocchio basis\cite{[Sto08]} or Woods-Saxon basis
\cite{[Sch08a],[Sch08b],[Zho03w],[Zho10],[Li12]} turned out to be an
efficient way to account for the continuum effects. Finally,
quasiparticle  continuum can be effectively discretized by solving
the HFB problem by means of expansion in a harmonic oscillator (HO)
or transformed HO (THO) basis\cite{[Sto00],[Sto03],[Sto05]}.  As far
as the description of nonlocalized HFB states is concerned, the
coordinate-space method is superior over the HO expansion method, as
the HO basis states are always localized. Consequently, the
discretized representation of the quasiparticle continuum is
different in coordinate-space and HO basis-expansion approaches
\cite{[Bor06]}.

Among the quasiparticle resonances, the deep-hole states play a
distinct role. In the absence of pairing, a deep-hole excitation with
energy $E_i>0$ corresponds to an occupied  HF state with energy
$\varepsilon_i=-E_i+\lambda$. If pairing is present, it  generates a coupling
of this  state with  unbound particle states with
$\varepsilon_i\approx E_i+\lambda$ that gives rise to a  quasiparticle
resonance with a finite width\cite{[Bel87],[Dob96],[Fay00a]}.
Quasiparticle resonance widths can be directly calculated with a high
precision using coordinate-space Green's function technique
\cite{[Bel87],[Oba09],[Zha11]} and  GHFB\cite{[Mic08]}. For
approaches based on the   $\mathcal{L}^2$ discretization, several
approximate methods have been developed to deal with HFB resonances.
The  modified stabilization method based on box solutions
\cite{[Man94a],[Zha08],[Pei11]} has been used to obtain
precisely the resonance energy and widths. Based on the box
solutions, the HFB resonances are expected to be localized solutions
with energies weakly affected by changes of the box size. The
stabilization method allows to obtain the resonance parameters from
the box-size dependence of quasiparticle eigenvalues.

Besides the
stabilization method, a straightforward smoothing and fitting method
that utilizes the density of box states has been successfully used.
In this technique, resonance parameters are obtained by fitting the
smoothed occupation numbers obtained from the dense spectrum
discretized HFB solutions.
\begin{figure}[htb]
\center
\includegraphics[width=0.45\textwidth]{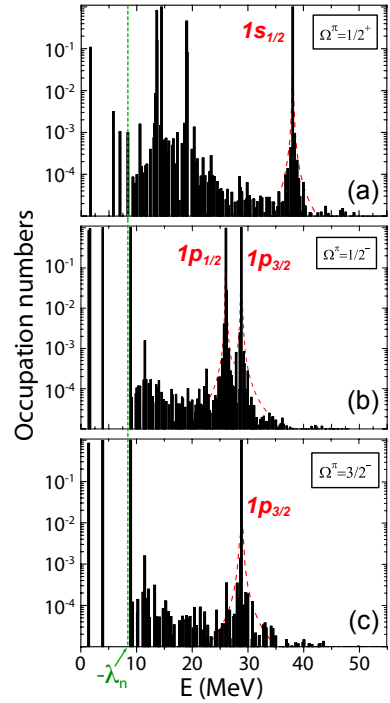}
\caption{\label{resonances} Occupation numbers of the discretized
neutron quasiparticle continuum states calculated for $^{70}$Zn in
Ref.\cite{[Pei11]}. The corresponding  Breit-Wigner envelopes are
indicated by dashed lines. The  $-\lambda_n$ threshold is marked by a
dotted line.}
\end{figure}
Figure~\ref{resonances} displays occupation probabilities $v_i^2$ for
the discretized neutron quasiparticle states in $^{70}$Zn as a
function of quasiparticle energy $E_i$. To extract resonance
parameters from the discrete distribution of $v_i^2$, one can  first
smooth it using a Lorentzian shape function and then perform a fit
using a Breit-Wigner function\cite{[Pei11]}.

Various ways of computing  widths of high-energy deep-hole states
have been compared in Ref.\cite{[Pei11]}. By comparing with the
exact GHFB results, it has been concluded that the stabilization
method works fairly well for all HFB resonances,  except for the very
narrow ones. The smoothing-fitting method is also very effective and
can easily be extended to the deformed case. The perturbative  Fermi
golden rule\cite{[Bel87]} has been found to be unreliable for
calculating widths of neutron resonances. (For more discussion on
limitations of the perturbative treatment, see Ref.\cite{[Bel87]}).

Pairing correlations can profoundly modify properties of the system
in drip line nuclei due to the presence of the vast continuum space
available for pair scattering. One example is the appearance of the
pairing-antihalo effect
\cite{[Ben99b],[Ben00],[Yam05a],[Rot09],[Rot09a],[Hag12]}, in which
pairing correlations in the weakly-bound even-particle  system change
the asymptotic behavior of particle density thus reducing its radial
extension. While  neutron radii of even-even nuclei  are expected to
locally increase when approaching the two-neutron drip line
\cite{[Miz00a],[Fay00a],[Sch08b],[Rot09],[Rot09a]} the size of the
resulting halo is fairly modest, especially when compared with
spatial extensions of neighboring odd-neutron systems.

Pairing  correlations  impact the limits of the nuclear existence:
the odd-even staggering of the nuclear binding energy does result in
the shift between one-neutron and two-neutron drip lines. The pairing
coupling to the positive-energy states is an additional factor
influencing the nuclear binding\cite{[Bel87],[Dob96]}. In
particular,  because of strong coupling to the neutron continuum, the
neutron chemical potential  may be  significantly lowered thus
extending the range of bound nuclei, and this effect is expected to
depend on the character of pairing interaction. For more discussion
on the impact of continuum on quasiparticle occupations, emergence of
bound canonical HFB states from the continuum, and contributions of
nonresonant   continuum to the localized ground state in dripline
nuclei, see
Refs.\cite{[Dob96],[Fay00a],[Rot09],[Rot09a],[Zha11],[Li12]}.

\section{Regularization of the local pairing interaction }\label{sec5}

\begin{figure}
\center
\includegraphics[width=0.60\textwidth]{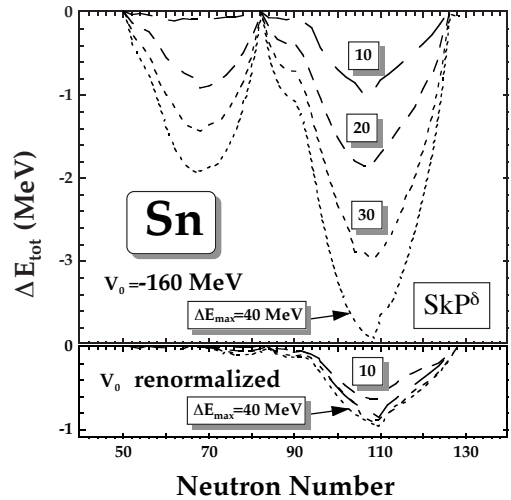}
\caption{Total energies in the tin isotopes calculated within the
HFB+SkP$^\delta$ model\cite{[Dob96]}.  Top panel shows the results
for the fixed interaction strength $V_0$ and for several cut-off
energies $\Delta E_{\text{max}}$ added to the usual
${\ell}j$-dependent cut-off energy $E_{\text{max}}$\cite{[Dob84]}.
Bottom panel shows similar results when the values of $V_0$ are
renormalized  to keep the average neutron pairing gap in $^{120}$Sn
the same for each $\Delta E_{\text{max}}$.}
\label{renorm}
\end{figure}

As discussed in Sec.~\ref{sec10}, in many HFB applications, pairing
interaction is often  assumed to be in the form of the zero-range,
density-dependent force. Calculations using the contact interaction
are numerically simpler, but the pairing gap diverges when the
dimension of the pairing-active space increases for a  fixed strength
of the interaction. In roots of this problem is the ultraviolet
divergence of abnormal density for zero-range pairing interaction
\cite{[Pap99],[Bul02],[Bor06]}:
\begin{equation}\label{divrho}
\left.\tilde\rho(\mb{r}-\mb{x}/2,\mb{r}+\mb{x}/2)\sim-\frac{\tilde h(\mb{r})
M^*(\mb{r})}{4\pi\hbar^2|\mb{x}|}\right|_{\mb{x}\rightarrow 0}.
\end{equation}
Consequently, in practical calculations,  one has to apply a cutoff
procedure to truncate the pairing-active space of s.p.\  states
\cite{[Dob84],[Dob96],[Rot09a]}, and the pairing strength has to be
readjusted accordingly. Thus the energy cutoff and the pairing
strength together define the pairing interaction, and this definition
can be understood as a phenomenological introduction of finite range
\cite{[Dob96],[Ber91],[Esb97]}. Such a sharp cut-off regularization
is performed in the spirit of the effective field theory, whereupon
contact interactions are used to describe low-energy phenomena while
the coupling constants are readjusted for any given energy cutoff to
account for  high energy effects. It has been shown that by an
appropriate renormalization the pairing strength for each value of
the cutoff energy, one practically eliminates the dependence of
various observables on the cutoff parameter\cite{[Dob96],[Bor06]}.
Figure{\ }\ref{renorm} illustrates the procedure for the total energy
in the tin isotopes. While for a fixed pairing strength total
energies depend significantly on
the cut-off energy (top), for a fixed pairing gap the
changes obtained with  renormalized interactions (bottom) are very
small indeed.

The cutoff energy dependence of the pairing strength can also be
handled by means of a regularization scheme by defining the
regularized local abnormal density
\cite{[Mar98],[Bru99],[Pap99],[Bul02],[Yu03],[Nik05],[Bor06]}:
\begin{equation}\label{rhoreg}
\tilde\rho_r(\mb{r})=\lim_{\mb{x}\rightarrow 0}
\left[\tilde\rho(\mb{r}-\mb{x}/2,\mb{r}+\mb{x}/2)-f(\mb{r},\mb{x})\right],
\end{equation}
where $f$ is a regularization counterterm, which removes the
divergence (\ref{divrho})  at $\mb{x}=0$. For cutoff energies high
enough, one can express $f$ through  the s.p.\  Green's function at
the Fermi level, $G(\mb{r}+\mb{x}/2,\mb{r}-\mb{x}/2)$, which also
exhibits a $1/x$ divergence. In practical calculations, one cane use
the Thomas-Fermi (TF) approximation for the local s.p.\  Green’s
function; this approach has been used with success for a description
of spherical and deformed nuclei
\cite{[Yu03],[Nik05],[Bor06],[Rot09a]}. As demonstrated in
Ref.\cite{[Bor06]}  the differences between pairing renormalization
and regularization procedures are rather small.

A combination of the renormalization and regularization methods
described above  is the hybrid technique\cite{[Pei11]} based on the
TF approximation to the non-resonant HFB continuum
\cite{[Rei99c],[Liu07]}. This approach is of great practical
interest as it makes it possible to carry out calculations in  wide
pairing windows  and very large coordinate spaces. In the hybrid
method, the high-energy continuum above the cutoff energy  $E_c$ is
divided into the non-resonant part and  deep-hole states. While
deep-hole states have to be treated separately, the non-resonant
continuum contribution to HFB densities and  fields can be integrated
out by means of the TF approximation. The choice of the cutoff $E_c$
is determined by positions of  deep-hole levels\cite{[Pei11]}; this
information can be obtained  by solving the HF problem.

\section{Pairing in odd-mass nuclei}\label{sec7}

The zero-quasiparticle HFB state~(\ref{eq:5}), representing the
lowest configuration for a system with even number of fermions,
corresponds to a filled sea of {\Bo} quasiholes with negative
quasiparticle energies, see Fig.~\ref{Sea}. In a one-quasiparticle state representing a
state in an odd nucleus, a positive-energy quasiparticle state
$\alpha$ is occupied and its conjugated quasihole partner is empty. The
corresponding wave function can be written as
\begin{equation}
\label{1qp}
    |\Phi\rangle^{(\alpha)}_{\text{odd}} = {\cal{}N} \alpha^+_\alpha
        \exp\left(\thalf\sum_{\nu\mu}
              Z^*_{\nu\mu} a^+_\nu a^+_\mu\right)\vac,
\end{equation}
where $\alpha^+_\alpha$ is the quasiparticle creation operator,
\begin{equation}
\label{eq3.1b}
           \alpha^+_\alpha = \sum_\nu  \left(  U_{\nu\alpha}a^+_\nu
                                           +   V_{\nu\alpha}  a_\nu \right),
\end{equation}
which depends on the quasiparticle state $\chi_\alpha$ (\ref{eq:9}).
Density matrix and pairing tensor of state (\ref{1qp}) can be obtained by exchanging in ${\cal{}U}$
columns corresponding to the quasiparticle and quasihole states,
$\chi_\alpha$ and $\phi_\alpha$. The corresponding
density matrix reads explicitly,
\begin{equation}\label{exactBlocking}
\rho_{\mu\nu}^{(\alpha)}   = \left( V^{*}V^{T} \right)_{\mu\nu}
+ U_{\mu\alpha}U^{*}_{\nu\alpha} - V^{*}_{\mu\alpha}V_{\nu\alpha},
\end{equation}
and similar holds for the pairing tensor. After the column
replacement, matrix $U^{(\alpha)}_{\mu\alpha'}$ of one-quasiparticle
state becomes singular and has null space of dimensions one. Hence,
the occupation number of one of the s.p.\  states equals
to 1. This fact is at the origin of the name ``blocked states''
attributed to one-quasiparticle states (\ref{1qp}). These states
contain fully occupied s.p.\  states that do not contribute
to pairing field\cite{[Ban74],[Fae80],[Dug01],[Ber09],[Per08]}.

The blocking can also be implemented, for some configurations, by
introducing two chemical potentials for different superfluid
components (two-Fermi level approach, 2FLA)\cite{[Sen07],[Bul08a]}
As demonstrated in Ref.\cite{[Ber09]}, such procedure is equivalent
to applying a one-body, time-odd field that   changes the
particle-number parity of the underlying quasiparticle vacuum. For
polarized Fermi systems, in which no additional degeneracy of
quasiparticle levels  is present beyond the Kramers degeneracy, the
2FLA is equivalent to one-dimensional, non-collective rotational
cranking.

When describing properties of odd-mass nuclei, one selects the lowest
quasiparticle excitations $E_\alpha$ and carries out the self-consistent
procedure based on these blocked candidates (\ref{1qp}). Naturally,
one must adopt a prescription to be able to determine, at each
iteration, the index $\alpha$ of the quasiparticle state to be blocked
\cite{[Hee05]}. Such a unique identification can be done  by means
of, e.g., the overlap method of  Ref.\cite{[Dob09d]}.  After the
HFB iterations are converged for each blocked candidate, the state
corresponding to the lowest energy is taken  as the ground state of
an odd-mass nucleus, and the remaining ones are approximations of the
excited states. A similar procedure can be applied to
many-quasiparticle states, e.g., two-quasiparticle states in
even-even and  odd-odd nuclei, three-quasiparticle excited states in
odd-mass nuclei, and so on.

The state (\ref{1qp}) represents an odd-Fermi system that  carries
nonzero angular momentum; hence, it breaks the time
reversal symmetry. If the time reversal symmetry is enforced,
additional approximations have to be applied based on  the Kramers
degeneracy. One of them is the equal filling approximation ({\EFA}),
in which the degenerate time-reversed states $\chi_\alpha$ and
$\chi_{\bar\alpha}$ are assumed to  enter the density matrix and
pairing tensor with the same weights\cite{[Per07],[Per08]}. For
instance, the blocked density matrix of {\EFA} reads:
\begin{equation}\label{EFA}
\rho_{\mu\nu}^{{(\alpha),\EFA}} = \left( V^{*}V^{T} \right)_{\mu\nu} +
\frac{1}{2}\left(
U_{\mu    \alpha}U^{*}_{\nu    \alpha} - V^{*}_{\mu    \alpha}V_{\nu    \alpha} +
U_{\mu\bar\alpha}U^{*}_{\nu\bar\alpha} - V^{*}_{\mu\bar\alpha}V_{\nu\bar\alpha}
\right).
\end{equation}
It has  been shown\cite{[Sch10]} that the {\EFA} and the exact
blocking are both strictly equivalent when the time-odd fields of the
energy density functional are put to zero. Thus, {\EFA} is adequate
in many practical applications that do not require high accuracy.

Although for the functionals restricted to time-even fields,  the
time-reversed quasiparticle states $\alpha$ and $\bar\alpha$ are
exactly degenerate, this does not hold in the general case. Here,
the blocking prescription may depend on which linear combination of
those states  is used to calculate the blocked density matrix. This
point can be illuminated by introducing  the notion of an {\it
alispin}\cite{[Sch10]}, which describes the  arbitrary unitary
mixing of $\chi_\alpha$ and $\chi_{\bar\alpha}$: $\chi'_\alpha =
a\chi_\alpha + b\chi_{\bar\alpha}$ ($|a|^{2} + |b|^{2} = 1$). As
usual, the group of such unitary mixings in a $2\times2$ space can be
understood as rotations of abstract spinors, which we here call
alirotations of alispinors. If the time-reversal symmetry is
conserved, the  blocked density matrix becomes independent of the
mixing coefficients $(a, b)$, that is, it is an aliscalar. In the
general case where time-reversal symmetry is not dynamically
conserved, however, the blocked density matrix is not aliscalar.
Here,  the blocked density matrix may depend on the choice of the
self-consistent symmetries and  the energy of the system may change
with alirotation.

The key point in this discussion  is the
realization that blocking must depend on the orientation of the
alignment vector with respect to the principal axes of the  mass
distribution. To determine the lowest energy for each quasiparticle
excitation,  self-consistent calculations should be carried out by
varying the orientation of the alignment vector with respect to the
principal axes of the system\cite{[Olb04],[Shi12]}. While in many
practical applications one  chooses a fixed direction of alignment
dictated by practical considerations, it is important to emphasize
that it is only by allowing the alignment vector to point out in an
arbitrary direction that the result of blocked calculations would not
depend on the choice of the basis used to describe the odd nucleus.
Illuminating examples presented in Ref.\cite{[Sch10]} demonstrate
that the choice of the alignment orientation does impact predicted
time-odd polarization energies.

Examples of self-consistent HFB calculations of one-quasiparticle
states can be found in
Refs.\cite{[Cwi99],[Afa03],[Sch10],[Afa11],[Bal12]} (full blocking)
and Refs.\cite{[Hil07],[Rod10],[Rod10a],[Li12a]} ({\EFA}).

\section{Summary and conclusions}\label{sec11}

The superfluid DFT based on self-consistent HFB  has already become the standard
tool to describe pairing correlations in atomic nuclei. Such framework has been
implemented in numerous approaches aiming at a consistent description of
particle-hole and particle-particle channels, and it is gradually replacing
a much simpler original BCS theory. This is so, because in
finite systems like nuclei, spatial dependence of particle and pairing fields has
to be properly described, especially in the nuclear periphery of
weakly bound isotopes. In this respect, the BCS theory and its different flavors are manifestly deficient\cite{[Bul80],[Dob84],[Bel87],[Dob96]}.

In this study, we aimed at presenting some basics of the local superfluid DFT  along with several aspects of it related to advanced current
applications. There are, of course,  numerous aspects of the HFB theory that we could
not cover in this limited overview. First, there have been  many
applications of the HFB theory using finite-range interactions, which
imply nonlocal pairing fields. While they are
significantly more difficult to treat, they do
not lead to ultraviolet divergencies. Based on the current description of the
limited set of nuclear observables related to pairing, it is difficult to judge whether the finite
range is essential. In fact, one can  understand  finite-range interactions in terms of regularized local functionals. Second, we did not
discuss various issues related to the restoration of  particle-number symmetry.
Effects of particle-number nonconservation are probably little
significant in heavy nuclei, but they may become crucial for some
observables and in specific systems, like, for example, nuclei with
only few particles in valence shells. Third, we could not cover
subjects related to the  treatment of pairing in  high-spin states where  the broken time-reversal symmetry
precludes the  use of the BCS theory.
Fourth, pairing correlations impact nuclear dynamics in a profound way. Recently, there have been many exciting developments related to the treatment of small- and large-amplitude collective motion in  weakly-bound superfluid nuclei.
Finally, we did not discuss details of the
HFB theory applied to the isoscalar pairing. This channel
becomes essential in nuclei with almost equal numbers of protons and
neutrons, and numerous applications of the HFB theory to this case
exist in the literature, see Refs.\cite{[Per04],[Roh10]} Many of these topics are discussed in other contributions contained in this Volume.\cite{(BroZel)}

\section*{Acknowledgments}
This work was supported in part by the
Academy of Finland and
University of Jyv\"askyl\"a within the FIDIPRO programme, and by the
U.S. Department of Energy under
Contract No.  DE-FG02-96ER40963.

\bibliographystyle{ws-rv-van}

\end{document}